\documentclass[conference]{IEEEtran}
\IEEEoverridecommandlockouts
\usepackage{amsmath,amssymb,amsfonts}
\usepackage{algorithmic}
\usepackage{graphicx}
\usepackage{textcomp}
\usepackage{xcolor}
\usepackage[nolist]{acronym}
\usepackage[binary-units]{siunitx}
\usepackage{comment}
\usepackage[
backend=biber,
style=ieee,
sorting=none
]{biblatex}
\def\BibTeX{{\rm B\kern-.05em{\sc i\kern-.025em b}\kern-.08em
    T\kern-.1667em\lower.7ex\hbox{E}\kern-.125emX}}

\begin{document}

\begin{acronym}
\acro{MD}{Molecular Dynamics}
\acro{AIMD}{Ab-Initio Molecular Dynamics}
\acro{FPGA}{Field Programmable Gate Array}
\acro{HPC}{High Performance Computing}

\acro{3d FFT}{three-dimensional Fast Fourier Transformation}
\acro{1d FFT}{one-dimensional Fast Fourier Transformation}
\acro{FFT}{Fast Fourier Transformation}
\acro{HPCC}{HPC Challenge}
\acro{OpenCL}{Open Computing Language}
\acro{GPU}{Graphics Processing Unit}
\acro{FLOPS}{Floating Point Operations per Second}
\acro{HDL}{Hardware Description Language}
\acro{AC}{Approximate Computing}
\acro{DSP}{digital signal processing}
\end{acronym}

\title{Efficient Ab-Initio Molecular Dynamic Simulations by Offloading Fast Fourier Transformations to FPGAs}

\author{
    \IEEEauthorblockN{
    Arjun Ramaswami\IEEEauthorrefmark{1},
    Tobias Kenter\IEEEauthorrefmark{2}, 
    Thomas D. Kühne\IEEEauthorrefmark{3} 
    and Christian Plessl\IEEEauthorrefmark{4
    }}
    \IEEEauthorblockA{
    Paderborn University, Paderborn, Germany\\
    \{\IEEEauthorrefmark{1}arjun.ramaswami,
    \IEEEauthorrefmark{2}kenter,
    \IEEEauthorrefmark{4}christian.plessl\}@uni-paderborn.de} \IEEEauthorrefmark{3}tkuehne@gmail.com
}

\maketitle

\begin{abstract}

A large share of today's \acs{HPC} workloads is used for \ac{AIMD} simulations, where the interatomic forces are computed on-the-fly by means of accurate electronic structure calculations. They are computationally intensive and thus constitute an interesting application class for energy-efficient hardware accelerators such as \acsp{FPGA}. 
In this paper, we investigate the potential of offloading 3D \acp{FFT} as a critical routine of plane-wave-based electronic structure calculations to \acs{FPGA} and in conjunction demonstrate the tolerance of these simulations to lower precision computations.
\color{black}
\end{abstract}

\section{Introduction}

\ac{AIMD} simulations are one of the most computationally intensive scientific simulations in current high performance cluster systems. These simulations involve computing interatomic forces using accurate electronic structure calculations 
to simulate the motion of atoms. Accelerating these simulations has led to massively parallel computations using multicore processors and hardware accelerators. Popular \ac{AIMD} simulation software packages such as Quantum Espresso~\cite{qepwscf} and CP2K~\cite{kuehne20_jcp} offer CUDA implementations to offload specific routines to GPUs. This scale of processing has led to maximizing performance through efficient computation.

One of the approaches to improve efficiency involves sacrificing accuracy in computations using lower precision arithmetic units, a strategy denoted as \ac{AC}. Techniques such as half-precision floating point arithmetic are used in hardware accelerators, specifically in recent Nvidia GPUs to support applications that tolerate approximation. \acp{FPGA} are becoming increasingly relevant in this regard, mainly due to the flexibility in building custom data paths and arbitrary widths/precision fixed or floating point units.

In \ac{AIMD}, a key kernel on the critical path of plane-wave-based density functional theory (DFT) calculations are 3D \acp{FFT}, which are exploited for an efficient evaluation of the quantum-mechanical operators.
Accelerating this routine involves optimizing execution in ranges of microseconds. In the context of classical force-field-based \ac{MD} simulations,~\citeauthor*{humphries20143d} implement high performance 3D \ac{FFT} with single precision floating point accuracy by fitting the entire data set in the \ac{FPGA} and by using vendor-specific 1D \ac{FFT} IP blocks. Their evaluation focuses on the performance of the 3D \ac{FFT} design, but does not evaluate the impact of reduced precision on the \ac{MD} application in question. This raises the question whether acceleration of specific routines with reduced precision can be tolerated by \ac{MD} simulations. Recently, \citeauthor*{rengaraj2020accurate} have shown that errors introduced by low precisions computing can be modeled as noise and can be perfectly corrected at the algorithmic level by means of a modified Langevin equation~\cite{CPMD}.

In this paper, the resilience of \ac{AIMD} simulations to approximation from lower precision floating point arithmetic is investigated by offloading 3D \ac{FFT} computations to \acp{FPGA}. This is demonstrated by designing an \ac{OpenCL} based 3D \ac{FFT} design for Intel \acp{FPGA} that uses single precision instead of double precision floating points for computations. This is compared with FFTW for performance and evaluated for \ac{AIMD} simulations by creating an interface in the CP2K framework. As an initial contribution, this interface has already been adopted as part of CP2K 7.1 release\footnote{https://github.com/cp2k/cp2k/releases/tag/v7.1.0}.

\section{Approach}
3D \ac{FFT} is an efficient algorithm to compute a 3D discrete Fourier transform (DFT) as defined in Equation~\eqref{eq:3dFFT}, where \( W_N = \exp(-i \frac{2\pi}{N}) \): 
\vspace{-0.1cm}
\begin{equation}
\label{eq:3dFFT}
    F( k_x , k_y , k_z ) = \sum_{z=0}^{N-1} \sum_{y=0}^{N-1} \sum_{x=0}^{N-1} 
    f(x,y,z) W_{x}^{xk_x} W_{y}^{yk_y} W_{z}^{zk_z}
\end{equation}

To compute the 3D \ac{FFT} of N\(^3\) points, the computation can be decomposed into 1D \acp{FFT} in \textit{x}-dimension followed by the same in the \textit{y}-dimension and then the \textit{z}-dimension.

\begin{figure}[h!]
    \centering
    \includegraphics[width=\linewidth]{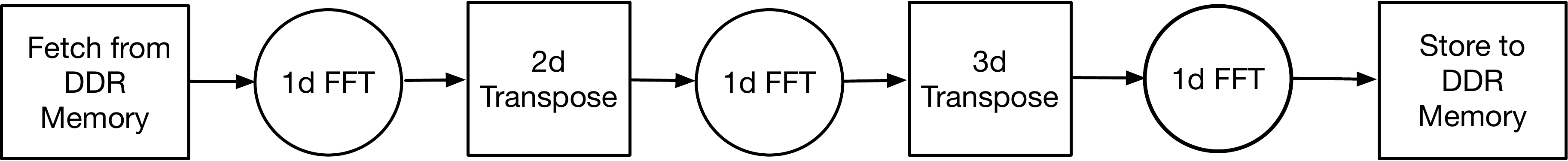}
    \caption{3D FFT Design for FPGAs}
    \label{fig:3dfft_design}
\end{figure}

The platform to offload this routine is the Nallatech 520N board that houses a Stratix 10 GX2800 \ac{FPGA} along with four 8 \si{\giga\byte} banks of DDR4 memory. It is connected to the host CPU using an 8-lane Gen3 PCIe bus. Developing an OpenCL 3D \ac{FFT} kernel design targeting this board requires transferring N\(^3\) points from the host CPU to the DDR (global) memory via the PCIe bus, transforming the data and finally, transferring the results back to the host CPU. The transformation step involves fetching data from the global memory followed by three N-point 1D \ac{FFT} computation phases intertwined by two transposition phases, the 2d matrix transposition and transposition along the z-dimension (3D transpose), and storing the result back to the global memory. The 1D \ac{FFT} computations are based on Intel's 1D \ac{FFT} \ac{OpenCL} design sample, which uses 8 single precision points per cycle in bit reversed order as input and output, transforming N points in $\frac{N}{8}$ cycles. The 2d transpose stage uses multiple buffers to avoid stalling the pipeline, however, this cannot be realized for the 3D transpose stage as it may not fit into the \ac{FPGA}. Figure~\ref{fig:3dfft_design} illustrates this design, where the entire 3D \ac{FFT} fits into the local memory, hence the 3D transpose is performed in the \ac{FPGA}. 

\vspace{-0.4cm}

\section{Results}

The 3D \ac{FFT} design described above is implemented for Intel's Stratix 10 \acp{FPGA}. It uses single precision floating point \ac{DSP} blocks in 1D \ac{FFT} computations. The code is available as an open source project\footnote{https://github.com/pc2/fft3d-fpga}. The \ac{FPGA} kernel code is synthesized using Intel OpenCL SDK 19.3 for Nallatech 520N board. The host code, which is compiled using GCC v8.3, measures the total wall clock time of PCIe data transfers between the host and the DDR memory as well as the time taken for kernel execution that includes \ac{FPGA} and DDR memory communication. This is measured over an average of a hundred iterations of non-batched executions.

Here, the performance of the Stratix 10 implementation is compared with the single precision floating point variant of FFTW v3.3.8. The FFTW application is linked using GCC v8.3 and executed on a node with two Intel Xeon Skylake Gold 6148 CPUs, each featuring 20 cores with hyperthreading disabled operating at 2.4 \si{\GHz}. Performance is measured by scaling the number of threads from 1 to 40 threads pinned to the 40 cores available for an average of hundred iterations each, using all four planning heuristics; the best runtime obtained among the different multithreaded executions is used for comparison, as shown in Table~\ref{tab:fft3d_runtime}. 

The performance of the \ac{FPGA} is promising considering the evaluation compares highly optimized libraries utilizing 40 core server CPUs. The latency of the \ac{FPGA} implementation can be reduced by overcoming the bottleneck with the 3D transpose step. PCIe data transfer latency can be overcome using larger transforms, making use of the available bandwidth. 


\begin{table}[]
  \centering
  \caption{Comparison of runtimes in milliseconds of different 3D FFT sizes between FFTW and FPGA 3D \ac{FFT} along with the latency of PCIe data transfers between host and DDR memory}
  \label{tab:fft3d_runtime}
  \begin{tabular}{|c|c|c|c|}
  \hline
  N$^3$ & FFTW & \begin{tabular}[c]{@{}c@{}}FPGA \\ Kernel Execution\end{tabular} & PCIe Data Transfer \\ \hline
  16$^3$ & 0.01 & 0.11 & 0.05 \\ \hline
  32$^3$ & 0.03 & 0.22 & 0.21 \\ \hline
  64$^3$ & 0.14 & 0.74 & 0.87 \\ \hline
  \end{tabular}
\vspace{-0.3cm}
\end{table}

In order to evaluate the resiliency of \ac{AIMD} simulations to approximate floating point arithmetic, an interface to the \ac{FPGA} 3D \ac{FFT} implementation is integrated into the CP2K framework. This is then compared with the default double precision CPU execution of 3D \ac{FFT} using FFTW3~\cite{frigo2005design}. The application chosen for evaluation uses the Gaussian and plane wave DFT method to compute the molecular orbitals of a single H$_2$O molecule in the gas phase~\cite{GPW}.
By restricting the plane-wave density cutoff, the application is evaluated for the specific 3D \ac{FFT} sizes that matches the \ac{FPGA} implementation. Both experiments converge to nearly identical total energies, leading to similar nuclear forces, thus showing that the \ac{AIMD} simulations are tolerant to approximations in floating point arithmetic.

With these initial accomplishments, further work can be focused on efficient acceleration of \ac{AIMD} simulations by developing a competitive 3D \ac{FFT} design for \acp{FPGA} that scales to larger \ac{FFT} sizes, where \acp{FPGA} should have a significant advantage. Moreover, investigating the tolerance of \ac{AIMD} simulations towards further approximation using custom precision floating point units in \acp{FPGA} could help achieve higher computational efficiency.  

\printbibliography

\end{document}